\documentclass[aps,reprint,nofootinbib,showpacs,amsmath,amssymb]{revtex4-1}
\usepackage{latexsym,graphicx,slashed,bm,dcolumn}

\begin{document}

\title{Fast Radio Bursts and Axion Miniclusters}

\author{Igor I. Tkachev}
\affiliation{Institute for Nuclear Research of the Russian Academy of Sciences, Moscow 117312, Russia}

\def\be{\begin{equation}}
\def\ee{\end{equation}}
\def\bea{\begin{eqnarray}}
\def\eea{\end{eqnarray}}
\def\la{\mathrel{\mathpalette\fun <}}
\def\ga{\mathrel{\mathpalette\fun >}}
\def\fun#1#2{\lower3.6pt\vbox{\baselineskip0pt\lineskip.9pt
        \ialign{$\mathsurround=0pt#1\hfill##\hfil$\crcr#2\crcr\sim\crcr}}}
\def\rhoeq{{{\rho_{\rm eq}}}}
\def\rhodm{{{\rho_{\rm DM}}}}
\def\Teq{{{T_{\rm eq}}}}
\def\aeq{{{a_{\rm eq}}}}
\def\bra{{{\bar{\rho}_a}}}

\begin{abstract}
Non-linear effects in the evolution of the axion field in the early
Universe may lead to the formation of gravitationally bound clumps of
axions, known as ``miniclusters.''  Minicluster masses should be in the range 
$M_{\rm mc}\sim10^{-12} M_\odot$, and in plausible early-Universe scenarios a 
significant fraction of the mass density of the Universe may be in the form of
axion miniclusters. Here I argue that observed properties (total energy release, duration, high brightness temperature, event rate) of recently discovered Fast Radio Bursts can be matched in a model which assumes explosive decay of axion miniclusters. 
\end{abstract}

\keywords{Fast Radio Bursts; Dark Matter --- axions}

\maketitle

\section{Introduction}

The recent detection of unusual radio pulses \cite{Lorimer:2007qn,Keane:2009cz,Thornton:2013iua,Spitler:2014fla,Burke-Spolaor:2014rqa}, known as Fast Radio Bursts (FRBs), has generated strong interest in identifying their origin and nature. The bursts exhibit a frequency-dependent time delay, which obeys a quadratic form so strictly, that the only explanation remains - signal dispersion in cold cosmic plasma during propagation. The magnitude of this delay, proportional to the electron column density along the line of sight, and called dispersion measure, is so large that the cosmological distances are inferred for the sources.

Thornton {\it et al}~\cite{Thornton:2013iua} deduce redshifts for four different FRBs observed at Parkes radio telescope to be in the range from 0.45 to 0.96. Recently detected FRB at Arecibo Observatory \cite{Spitler:2014fla} (which has larger size antenna and therefore excludes atmospheric artefacts as possibility for FRB) has derived redshift of $z=0.26$. All observers agree on a high rate of FRBs, $\sim 10^4$ events/day for the  whole sky.

FRBs are also characterized by extremely high flux densities ($\sim$ Jy) over very short time scales (milliseconds). 
Short time scales imply that the size of emitting region is small, less then $300$ km.
Observed fluxes imply that the total energy radiated in the band of observation was in the range $10^{38}-10^{40}$ ergs \cite{Thornton:2013iua,Spitler:2014fla}, assuming isotropy and quoted redshifts. Derived redshifts, and therefore the radiated energy, can be smaller if significant part of dispersion measure accumulates in the host galaxies. 
With this parameters, and assuming FRBs are at Gpc distances, their brightness temperature would be $T_B \sim 10^{36} ~{\rm K}$, leading to the conclusion that radiation from FRB sources should be coherent \cite{Katz:2013ica,Luan:2014iea,Kulkarni:2014vea,Pietka:2014wra}.

The source of the FRB signals is hotly debated in the literature, with suggested progenitors ranging from terrestrial interference to neutron star-neutron star mergers. Wide range of models was thoroughly discussed in Kulkarni {\it et al.} \cite{Kulkarni:2014vea}, and refs. therein. While all models have problems, the 
verdict was that several arguments \cite{Popov:2007uv}, which relate giant flares of young magnetars with FRBs, may offer  plausible physical scenario \cite{Lyubarsky:2014jta} based on assumption that FRBs could be attributed to synchrotron maser emission from relativistic, magnetized shocks.

FRBs are so mysterious that a new physics models were also suggested and discussed in literature \cite{Vachaspati:2008su}-\cite{Iwazaki:2014wka}.
E.g., Ref.~\cite{Luan:2014iea} even discussed, {\it en route}, the possibility that FRBs are signals beamed at Earth by advanced civilizations. 

In this paper I consider possible relation of FRBs and axions~\cite{Agashe:2014kda}.
In Ref.~\cite{Tkachev:1986tr} it was suggested that axion field may form gravitationally bound compact astrophysical objects, where under some conditions parametric instability occurs, resulting in a powerful coherent burst of maser radiation. Such instability has been also studied in Refs.~\cite{Kephart:1994uy,Riotto:2000kh}. Here I reconsider this scenario and discuss several mechanisms where axion miniclusters \cite{Kolb:1993zz,Kolb:1993hw,Kolb:1994fi,Kolb:1995bu} are responsible for Fast Radio Bursts, see also~\cite{Iwazaki:2014wka} . I'd like to note, that a general case of axion like particles (ALP, where particle mass, self coupling, and coupling to electromagnetic field are not tightly related to each other), opens a lot of possibilities which are ruled out otherwise. I do not consider general case of ALP, staying instead with the standard QCD invisible axion model. Generalization to ALP is straightforward.

\section{Dense Axion Objects}

The invisible axion is among the best motivated candidates for cosmic
dark matter~\cite{Agashe:2014kda}. The axion is the pseudo-Nambu--Goldstone boson resulting from the spontaneous breaking
of a $U(1)$ global symmetry known as the Peccei--Quinn, or PQ,
symmetry introduced to explain the apparent smallness of strong
CP-violation in QCD \cite{Peccei:1977hh}.

There are stringent astrophysical, cosmological, and laboratory constraints on the
properties of the axion~\cite{Agashe:2014kda}. In particular, the combination of cosmological and astrophysical considerations restricts 
the axion mass $m_a$ to be in the window ${\mu\rm eV} \alt
m_a \alt {\rm m eV}$. Corresponding value of the axion decay constant $f_a$ can be found using relation $f_a m_a = f_\pi m_\pi$, where $\pi$ referes to pion. The contribution to the mean density of the Universe
from axions in this window is guaranteed to be cosmologically
significant.  Thus, if axions exist, they will be dynamically
important in the present evolution of the Universe.

For what follows, it is important for PQ-symmetry to be restored after inflationary stage of the Universe evolution. This happens if reheating temperature is larger than the corresponding PQ-scale, but this may also happen~\cite{Kofman:1995fi} already during reheating, in a transient highly non-equilibrium state, even if resulting  temperature is small. In this situation the axion field takes different values in different casually disconnected regions at temperatures well above the QCD confinement, when the axion is effectively massless. At the confinement temperature and below,
QCD effects produce a potential for the axion of the form
$V(\theta)=m_a^2f_a^2[1-\cos(\theta)]$, where the axion field was parametrized
as a dimensionless angular variable $\theta \equiv a/f_a$. Axion oscillations commence and field variations are transformed into density contrasts, $\rho_a$, which
later lead to tiny gravitationally bound ``miniclusters''.

\subsection{Axion Miniclusters}

Since density variations in this circumstances are large from the very beginning, we do not refer to them as “perturbations”, and let us call corresponding regions as "clumps". It is easy to understand that today those clumps, or miniclusters, will be very dense objects. Let us specify the density of a dark-matter clump prior to matter-radiation equality as $\delta\rho_a/\rho_a \equiv \Phi$. In situation when $\Phi \sim 1$ (which would arise for non-interacting field $V(\theta)=m_a^2f_a^2\theta^2$ with random initial conditions), these clumps separate from cosmological expansion and form gravitationally bound objects already at $T=\Teq$, where $\Teq$ is the temperature of equal matter and radiation energy
densities. Density of such clump  today will correspond to a matter density back then, i.e. will be $10^{10}$ times larger than the local galactic halo dark matter density.

However, at the time when axion oscillations commence, in many regions $\theta \sim 1$, and self-interaction is important. Numerical investigation of the dynamics of the axion field around the QCD epoch \cite{Kolb:1993hw,Kolb:1993hw,Kolb:1994fi,Kolb:1995bu} had shown that the non-linear effects result in regions with $\Phi$ much larger than unity, possibly as large as several hundred, leading to enormous minicluster densities.
In such situation a clump separates from cosmological expansion at $T \simeq (1+\Phi)\Teq$ which leads to a final minicluster density today given by \cite{Kolb:1994fi} 
\begin{equation}
\label{rhofl}
\rho_{\rm mc} \simeq 140 \Phi^3 (1+\Phi) \bar{\rho}_a(T_{\rm eq}).
\end{equation}
Even a relatively small increase in $\Phi$ is important because the final
density depends upon $\Phi^4$ for $\Phi\ga 1$. 

The scale of minicluster masses is set by the total mass in axions within the
Hubble radius at a temperature around $T \approx 1$ GeV when axion
oscillations commence, which is about $10^{-12}M_\odot$.
Masses of miniclusters are relatively insensitive to the particular
value of $\Phi$ associated with the minicluster.
Corresponding minicluster radius as a function of $M$ and $\Phi$:
\begin{equation}
R_{\rm mc} \approx  \frac{2 \times 10^{6}}{\Phi \left(1+\Phi\right)^{1/3}
\Omega_a h^2} \left(\frac{M}{10^{-12} M_\odot}\right)
^{1/3} {\rm km} \, .
\label{R}
\end{equation}
Since large-$\Phi$ miniclusters are very dense, form
early, and are well separated from each other, they should escape
tidal disruption and merging.

According to Ref.~\cite{Kolb:1995bu}, more than 13\% of all axionic dark matter are in miniclusters with $\Phi \ga 10$,
more than about 20\% are in miniclusters with $\Phi
\ga 5$  and 70\% are in miniclusters $(\Phi > 1)$. Since roughly half of all axions reside in miniclusters, the total number of miniclusters in the Galaxy is large, 
$N \sim 10^{24}$.

At this point, an important relation
\begin{equation}
M_{\rm mc} \sim 10^{-12}M_\odot = 2\times 10^{42}\; {\rm ergs},
\end{equation}
and the fact that gigahertz frequency radiation is within allowed axion mass range, 
$\nu = {m_a}/{2\pi} \approx 2.4\, (m_a/10\, {\mu\rm eV})$ GHz, should tell us that if a fraction of axion minicluster mass is rapidly transformed into radiation, this will lead to something similar to observed FRB. 
Interestingly, in my notes dating back to 1997 I have found the following phrase: ``Even if the tiny fraction $10^{-25}M_\odot$ of the minicluster mass $10^{-12}M_\odot$ will go into radiation on this frequency, it can be detected from anywhere in the Galaxy halo (L. Rosenberg, private communication)''.  Back then this (and several uncertainness which I will describe below) actually had prevented me from submitting already prepared paper.
Let us discuss farther evolution of axion miniclusters and possible mechanisms of their mass transfer into radiation.

\subsection{Axion Bose-clusters}

Miniclusters with $\Phi \ga 30$ undergo the Bose-condensation later on and consequently became even denser and more compact \cite{Kolb:1993zz}. Usually, in the related literature, an existence of a Bose-star is just postulated, without questioning of how it can be formed, for a review of Bose-stars see, e.g.~\cite{Schunck:2003kk}. However, in the case of invisible axion all couplings are so small, that mere possibility of condensate formation has to be studied \cite{Tkachev:1991ka,Semikoz:1994zp,Khlebnikov:1999qy,Sikivie:2009qn}. Simple estimates has been done in Ref.~\cite{Tkachev:1991ka}, while Bose-condensation in the frameworks of Boltzmann equation was studied numerically in Ref.~\cite{Semikoz:1994zp}.
In Boltzmann approach Bose-condensate does not form actually, one can see only an establishment of the Kolmogorov inverse cascade towards zero momenta (but this approach does allow to estimate formation time). The problem was solved in~\cite{Khlebnikov:1999qy} by studying numerically the evolution of initially random classical fields, both for positive and negative self-couplings (the later corresponds to the axion case).

It is remarkable that in spite of the apparent smallness of axion quartic
self-coupling, $|\lambda_a| \approx (f_\pi/f_a)^4 \sim 10^{-53}
f_{12}^{-4}$, the subsequent relaxation in an axion minicluster
due to $2a \rightarrow 2a$ scattering can be significant as a
consequence of the huge mean phase-space density of axions. Then, 
instead of the classical expression,
$t_R^{-1} \sim \sigma \rho_a v_e m_a^{-1}$, where
$\sigma$ is the corresponding cross section and $v_e$ typical velocity in the gravitational well, one gets~\cite{Tkachev:1991ka} for the relaxation time
\begin{equation}
t_R^{-1}  \sim  \lambda_a^{2}\rho_a ^{2}v_e^{-2} m_a^{-7} \, .
\label{rl}
\end{equation}

The relaxation time (\ref{rl}) is smaller then the present age
of the Universe for miniclusters with $\Phi \ga 30$ \cite{Kolb:1993zz}. I will call  resulting objects Bose-clusters, not Bose-stars, since their total mass is not in a stellar mass range.

Characteristic sizes and limiting masses of resulting objects  can be estimated~\cite{Tkachev:1986tr}
analyzing simple equation of ``hydrostatic equilibrium'' in non-relativistic limit \begin{equation}
{dP(r) \over dr} = -{\rho (r) M(r) \over M_{\rm Pl}^2 r^2} \, .
\label{he}
\end{equation}
The pressure $P(r)$ and density $\rho(r)$ 
has to be understood here as quantities averaged over the period of field 
oscillation. 

{\it 1. Non-interacting field.} 
Positive contribution to the pressure comes 
from the field gradients, and can be approximated as $P_{\rm grad} 
\sim a_0^2/R^2$, where $R$ is the averaged size of the configuration, and subscript ``0'' in what follows will mean amplitude of field oscillations.
This gives
\begin{equation}
R \approx \frac{1}{m_a v_e} \approx 300\; \frac{10^{-12}M_\odot}{M_{bc}}\left(\frac{10\, {\mu\rm eV}}{m_a}\right)^2\; \rm km,  
\end{equation}
which, depending upon axion parameters and mass of the cluster can be comparable or less then   FRB's emitting region. The maximum possible mass of a stable Bose-cluster corresponds to $v_e \sim 1$ which gives
$M_{\rm max}({\lambda =0}) \approx M_{\rm Pl}^2 /m_a$. For non-interacting 
axions this would be in the range of $\sim 10^{-5} M_\odot$.

{\it 2. Positive self-coupling.}
The self-coupling may be tiny, but its contribution
to the pressure, $P_\lambda \sim \lambda a^4$, can not be neglected in a 
certain parameter range. Using this expression
in Eq. (\ref{he}) one finds~\cite{Tkachev:1986tr}
$M_{\rm max}({\lambda > 0}) = \sqrt{\lambda} M_{\rm Pl}^3 /m^2$. With the 
positive self-coupling (which corresponds to a repulsive interaction) the maximum 
mass of a stable Bose-cluster can be significantly bigger than for non-interacting particles. 

{\it 2. Negative self-coupling.}
The self-coupling of axions is negative and their interaction is attractive.
Consequently, there will be no-stable configuration when
$|P_\lambda| > P_{\rm grad}$.  We find that the instability develops when 
$|\lambda | {a_0}^2 R^2  > 1$ or at $M \agt M_{\rm Pl}/\sqrt{|\lambda |}$. For axions 
$|\lambda |  = m_a^2/f_a^2$ and this condition reduces to
$M_{\rm max}({\lambda < 0}) = f_a \,M_{\rm Pl}/m_a  \sim 10^{-12} M_\odot\, (10\, {\mu\rm eV}/m_a)^2$. Instability condition can be also re-written as 
$\theta_0 > v_e$.

During  Bose-relaxation the mass of the Bose-condensed core in the clump 
grows, while its radius shrinks. When the mass exceeds $M_{\rm max}({\lambda < 0})$, the core collapses. At this moment its radius is equal to 
\begin{equation}
R_{\rm min}  \sim M_{\rm Pl} /f_a m_a \approx 200~ \rm km, 
\end{equation}
regardless of $m_a$. 
Note that the maximum mass for a stable axion Bose-cluster  at $m_a = 10\, {\mu\rm eV}$ is of the order of the typical mass of the axion minicluster.

\section{Transfer of energy into radiation}

Electromagnetic properties of axions are described by Maxwell's equations
\begin{eqnarray}
\label{ame}
& \nabla & [{\bf E} + \frac{\alpha}{2\pi}\theta {\bf B}] =0\, ,  \\
& \nabla &\times [{\bf B} - \frac{\alpha}{2\pi} \theta {\bf E}] -
\partial_0[{\bf E} + \frac{\alpha}{2\pi}\theta {\bf B}]=0 \, , \\
& \nabla &{\bf B} =0, \,\,\,\,\,\,\,\,
\nabla \times {\bf E} + \partial_0{\bf B}=0 \, ,
\label{bme}
\end{eqnarray}
where $\alpha \approx 1/137$. There are several mechanisms by which considerable fraction of axion minicluster can be transferred into electromagnetic radiation, either for isolated cluster, or for cluster in magnetic field. Let us discuss isolated cluster first, for which stimulated decays $a \rightarrow \gamma \gamma$ were studied in Refs.~\cite{Tkachev:1986tr,Kephart:1994uy,Riotto:2000kh,Kolb:1993zz}.

{\it 1. Explosive maser effect.} 
In homogeneous axionic medium the Fourier amplitudes, $g_k$, of left and right- polarized photons will obey the equation 
\begin{equation}
\ddot{g}_k + (k^2 \pm k\alpha \dot{\theta}) g_k =0 \, ,
\label{math}
\end{equation}

This equation can be reduced to the standard form for the 
Mathieu equation $\ddot{g}_k + [A -2q \cos (2\tau)] g_k =0$
with $A \equiv 4k^2/m^2$ and $q \equiv 2k\alpha \theta_0/m$.
The number of photons in certain momentum bands will grow exponentially in time, 
$n_k =\exp (\mu_k t)$.
Maximum amplification is achieved for $k =m/2$ with $\mu=\alpha \theta_0 m/2$
(which corresponds to axion decay $a \rightarrow \gamma \gamma$),
and radiation is amplified in the band $\delta k =\mu$.
When incident radiation pass through the cluster, it will be amplified
along the photon path.
It is important that the position of the resonance, $k =m/2$, does not
depend upon the field amplitude, and only the $\mu$ does.

It is convenient to introduce the amplification coefficient for the whole
cluster, $D \equiv \mu R$. If at some moment of time the condition $D \gg 1$ is reached, the cluster will explode~\cite{Tkachev:1986tr}. We find $D \sim \alpha {\theta_0} /2 v_e$
for the equilibrium Bose-cluster, $R \sim 1/mv_e$.
In the axion case the cluster is in "hydrostatic" equilibrium when 
$\theta_0 < v_e$, therefore $D < \alpha/2 $. While clusters with $D < 1$ do not explode, axion decay width with respect to stimulated emission is much larger there as compared to free axions. Their decay may make an excess radio background. It is intriguing to note in this respect that  the extragalactic radio background measured by ARCADE~2~\cite{Fixsen} is a factor of $\approx 5$ brighter than the estimated contribution of resolved point sources. This excess was not explained.

Axionic Bose-cluster can explode only after it has lost equilibrium and is collapsing. 
What happens next is a difficult problem, requiring numerical modeling. I can only note that during collapse the condition $\theta_0 \sim 1$ can be reached. In this regime the system is not in a "narrow" resonance regime anymore (though not in a "wide"  regime either, where explosive decay can happen regardless of time-evolving background~\cite{Kofman:1994rk}). Also, quasi-stable configuration resembling "breezers" or "oscillons" may form~\cite{Kolb:1993hw} during collapce, which may help resonance to develop. Finally, in this regime the axion field does not evolve independently, but mixes with the pion field,
$\pi^0$. Indeed, physical pion and axion are mass eigenstates of
small oscillations around the minima of the common potential 
(see e.g. \cite{Huang:1985tt})
\begin{equation}
V(\theta,\beta)=f_\pi^2 m_\pi^2\, [\cos \theta \cos \beta -\xi \sin \theta
\sin \beta +1] \, ,
\label{ax-pion}
\end{equation}
where $\xi \equiv (m_d-m_u)/(m_d+m_u) \approx 0.3 $ and $\beta \equiv 
\pi^0/f_\pi $. With $\theta \sim 1$, the pion field will develop too. Which particle species and with what spectra will be created, and what would be the rate of corresponding explosions are interesting problems to study. 

{\it 2. Decay in magnetic field.} In strong magnetic fields axions can oscillate into photons. These oscillations are employed in axion laboratory searches and may happen in astrophysical environment too~\cite{Sikivie:1983ip}.

Resonant conversion happens when the photon plasma mass equals to the axion mass. In changing magnetic fields and electron plasma density this condition may be met at some distance from the astrophysical object.  I will not question all range of posible astrophysical environments,  but simply use resent work \cite{Pshirkov:2007st} where resonant conversion of dark matter axions to photons in magnetospheres of neutron stars has been considered. 

Calculations were presented for a particular example of the neutron star with magnitude of magnetic field on its surface $10^{14}$~G, and spin period 10 s. It was found, that conversion probability reaches $P = 0.2$  for axions with  $m_a = 5~ \mu eV$ at a distance $r = 3.4\, r_{ns}$, where magnetic field equals $2.5 \times 10^{12}$~G. Conversion reaches maximum possible value of $P = 0.5$ for $m_a = 7~ \mu eV$, and stays at this value for larger axion masses. Bandwidth of the signal was found to be in the range of $5$ MHz, however, calculations in Ref.~\cite{Pshirkov:2007st} where done for the case of unclustered halo axions. In the case of explosive conversion of the (fraction) of minicluster,  with all accompanying plasma effects the bandwidth will be higher. E.g., in Ref.~\cite{Luan:2014iea} it was noted that  electric fields near FRBs which are at cosmological distances would be so strong that they could accelerate free electrons from rest to relativistic energies in a single wave period. Questions related to resulting spectrum, burst duration and energy release require further detailed study.

Similarly to the case of a free minicluster, stimulated axion-photon conversion is possible in external magnetic field as well. To find the rate of stimulated emission one has simply to solve classical field equations, see Ref.~\cite{Khlebnikov:1996mc}, which in the present case will be  Eq.~(\ref{ame})-Eq.~(\ref{bme}). Axion minicuster is not a collection of a particles, but, as a consequence of a huge phase-space density, can be described as a random classical field, therefore this approach is valid.  
(Eq.~(\ref{ame}) in external magnetic field suggests particular solution
${\bf E} = -  {\alpha}\theta {\bf B_{\rm ext}}/{2\pi}$, but this can be misleading since the source for electric field is actually zero in homogeneous axion field. The whole system of equations has to be solved.)
When stimulated process develop in full strength, such that back reaction is important, resulting spectra of created particles are not narrow, but display series of peaks \cite{Khlebnikov:1996mc}, which evolve later on \cite{Micha:2002ey} into power law spectra of Kolmogorov turbulence.

\begin{figure}
\includegraphics[width=0.9\linewidth]{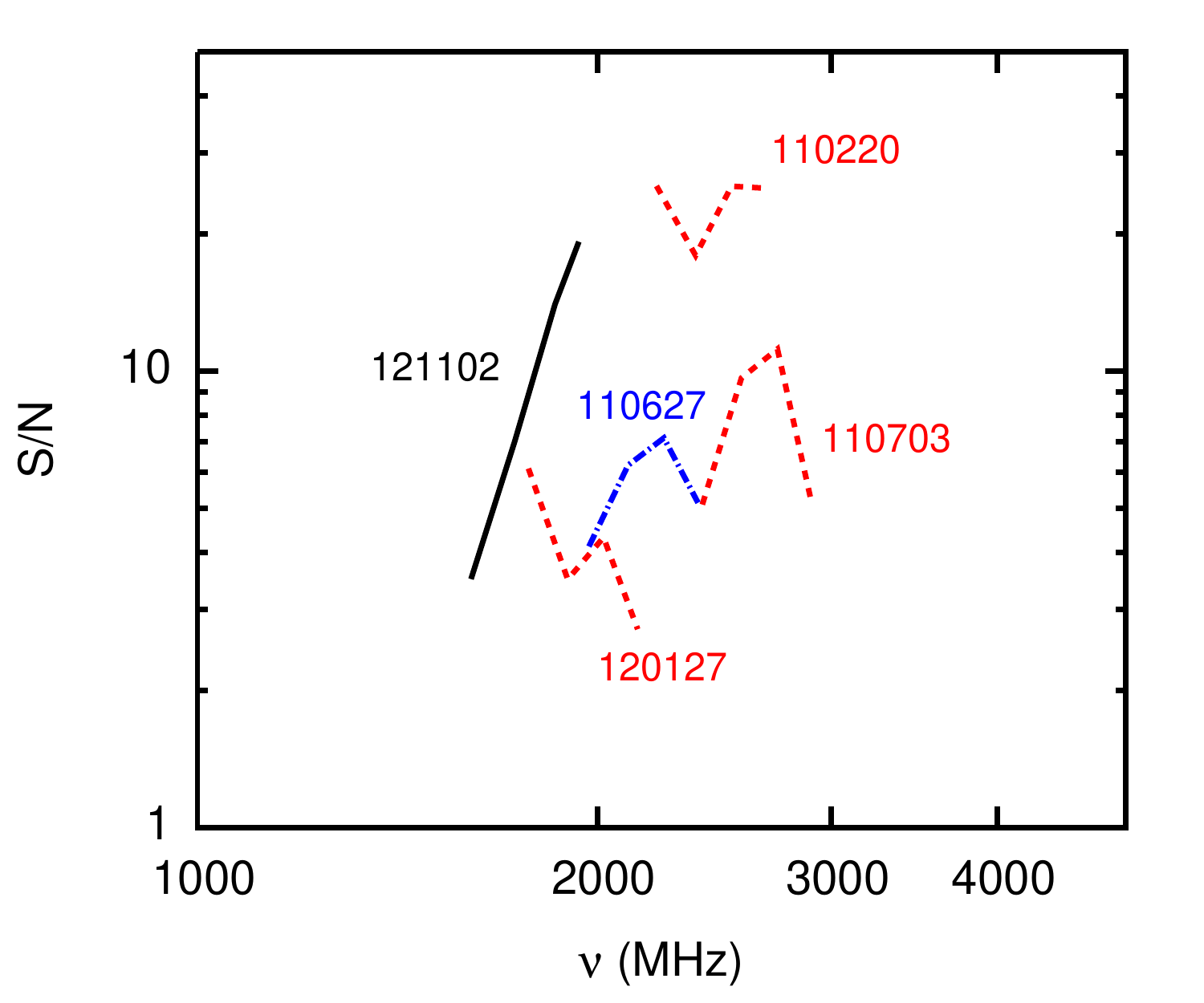}
\caption{Signal to noise ratios for detected FRB's after frequency shift by cosmological redshifts, $(1+z$). Solid line corresponds to data from Ref.~\cite{Spitler:2014fla}, while dashed lines to Ref.\cite{Thornton:2013iua}.}
\label{fig:spectra}
\end{figure}

Let us estimate event rate in this case.
Number density of miniclusters is $n \approx 10^{10}~\rm pc^{-3}$. Gravitational capture radius for a neutron star in a halo with typical velocities $v_h \sim 10^{-3}$ gives for 
cross-section $\sigma \approx 5\times 10^{13}~ \rm km^2$. This gives for event rate $n\sigma v_h \sim 4\times 10^{-10}~\rm day^{-1}$ for a collisions with single neutron star. Taking $10^{5}$ as an estimate for a number of neutron stars with strong magnetic field, and multiplying by the number of galaxies in a visible Universe we obtain $4\times 10^{6}~\rm day^{-1}$ for the event rate in this scenario.

\section{Discussion}

We have shown that observed properties of FRB's (total energy release, duration, high brightness temperature, event rate) can be matched in a model which assumes explosive decay of axion miniclusters. Primary frequency of radiation will correspond to $m_a/2$ for the decay of isolated minicluster, and to $m_a$ for the decay in external magnetic field. Comparison  of predicted and observed spectra may help to falsify the model. In particular, the (narrow) frequency band of resulting signal, in the reference frame of the source, should be always  at the same position in suggested model, while in pure astrophysical scenarios, e.g. in the model of Ref.~\cite{Lyubarsky:2014jta}, there is no particular reason for this to hold.  

In Fig.~(\ref{fig:spectra}) the signal to noise ratios for several FRBs observed in Refs.~\cite{Thornton:2013iua,Spitler:2014fla}  are presented. With respect to the original data, I have made a frequency shift by $(1+z)$, using cosmological redshifts quoted for each FRB. The Aresibo burst, shown by the solid line, has unusual (as it was noted in Ref.~\cite{Spitler:2014fla}), steeply rising spectrum, $S \propto \nu^\alpha$, with the best fit value $\alpha = 11$. This peculiarity of the signal was interpreted as a consequence of being detected in a sidelobe of the receiver. For the highest S/N event in Ref.~\cite{Thornton:2013iua} another peculliary has been stressed: spectrum has well defined bands, of 100 MHz width. For the other three events, the verdict was that they do not have sufficiently high S/N to say something definite about their spectra.
All of the above can be true. On the other hand, as Fig.~\ref{fig:spectra} suggests, the S/N for all of this bursts repeat similar pattern, which may be actually due to a narrow bandwidth of a maser emission. Visible, rather narrow peaks are not at the same frequency. However, spectra of well developed stimulated emission  may display several peaks~\cite{Khlebnikov:1996mc}. Also, redshifts to FRBs are not really known because of a possible significant dispersion at the sources.

\acknowledgments

I am grateful to  S. Popov, K. Postnov, M. Pshirkov, and S. Troitsky for useful discussions and comments.
This work was supported by the Russian Science Foundation grant 14-22-00161.

\end{document}